\documentclass[twocolumn]{aa}
\usepackage{graphicx}
\usepackage{graphics}
\usepackage{psfig}
\usepackage{natbib}  
\begin{document}
\bibliographystyle{bibtex/aa} 
\title{Probing the cosmic microwave background temperature 
using the Sunyaev-Zeldovich effect}
\author{
Cathy Horellou, 
Martin Nord, 
Daniel Johansson,
Anna L\'evy
          }

   \offprints{horellou@oso.chalmers.se}

   \institute{Onsala Space Observatory, 
	Chalmers University of Technology, 
  	SE-439 92 Onsala, Sweden}

   \date{Received 21 March 2005;}

   \abstract{ 
We discuss the possibility to constrain the relation between 
redshift and temperature of the cosmic microwave background (CMB) 
using multifrequency Sunyaev-Zeldovich (SZ) observations.  
We have simulated a catalog of clusters of galaxies detected through their
SZ signature assuming the  
sensitivities that will be achieved by the {\it Planck} satellite at 100, 143 and 
353~GHz, 
taking into account the instrumental noise and the contamination from 
the Cosmic Infrared Background and from unresolved radiosources.
We have parametrized the cosmological temperature-redshift law as $T\propto (1+z)^{(1-a)}$.
Using two sets of SZ flux density ratios 
(100/143~GHz, which is most sensitive to the parametrization of 
the $T-z$ law, and 143/353~GHz, which is most sensitive to the 
peculiar velocities of the clusters)
we show that it is possible to recover the $T-z$ law
assuming that the temperatures and redshifts of the clusters are known. 
From a simulated catalog of $\sim 1200$ clusters, 
the parameter $a$  can be recovered 
to an accuracy of 10$^{-2}$. 
Sensitive SZ observations thus appear as a potentially useful tool to test the standard law. 
Most cosmological models predict a linear variation of the
CMB temperature with redshift. 
The discovery of an alternative law would have profound implications 
on the cosmological model, implying creation of energy in a manner that
would still maintain the black-body shape of the CMB spectrum at 
redshift zero. 
   }
\titlerunning{Redshift dependence of the CMB temperature}
\authorrunning{Horellou et al.}
\maketitle
\section{Introduction} 

One of the strongest predictions of the standard cosmological model is 
the linear variation of the Cosmic Microwave Background (CMB) temperature 
with redshift:
$T_{\rm CMB}(z)=T_0 (1+z)$ where 
$T_0$ is the temperature of the CMB at redshift zero.
$T_0$ has been determined with a high accuracy from data taken   
by the Far Infrared Absolute Spectrometer (FIRAS) onboard the COBE satellite: 
$T_0=2.725\pm0.002$~K (\citealt{1999ApJ...512..511M}). 
Those measurements do not imply that the CMB spectrum is Planckian at all redshifts,
nor give information about how the temperature varies with redshift. 
Alternative cosmological models imply a different redshift dependence of the CMB temperature, 
for instance: 
\begin{equation}
T=T_0 (1+z)^{(1-a)}$ with $0 \leq a\leq 1 
\label{eq1}
\end{equation}
(\citealt{2000MNRAS.312..747L}). 
Equation~(\ref{eq1}) requires   
energy injection into the CMB, which could come from decay of vacuum energy 
(\citealt{1987NuPhB.287..797F}). 
In such a case, the Planckian form of the radiation spectrum is preserved if the creation 
of photons is adiabatic, in the sense that the entropy per photon remains constant 
(\citealt{1996PhRvD..54.2571L}). 

An important test of such models would be to determine the temperature of the CMB
at different redshifts through observations. 

Shortly after the discovery of the CMB by \cite{1965ApJ...142..419P} 
it was realized that the relic radiation had already been indirectly observed  
through observations of interstellar absorption lines with very low excitation temperatures  
(see \citealt{1972ARA&A..10..305T} for a review of early observations).  
More recently, a temperature $T_0=2.729^{+0.023}_{-0.031}$K
has been derived 
from a survey to determine the amount of CN rotational excitation 
in interstellar clouds  
(\citealt{1993ApJ...413L..67R}). 
\cite{1968ApJ...152..701B} 
pointed out that the temperature of the CMB could be probed at much higher
redshifts by studying the absorption spectra of quasars and searching for the species 
with energy states whose populations are sensitive to the ambient radiation temperature. 
This method, however, gives only an upper limit on the CMB temperature 
unless the local contribution to the excitation can be estimated, 
since other processes can contribute to the excitation of the atoms/molecules  
(\citealt{1997ApJ...474...67G}, 
\citealt{2000Natur.408..931S},  
\citealt{2002A&A...381L..64M}). 
So far, the standard law is consistent with observations, but so are  
alternative laws with $a= 0.003\pm0.13$,  
or $b= 0.99\pm0.22$
for a straight line $T(z)=T_0(1+bz)$
(\citealt{2001PhRvD..64l3002L}; see also \citealt{2004A&A...422....1P}). 

\begin{figure}
\psfig{file=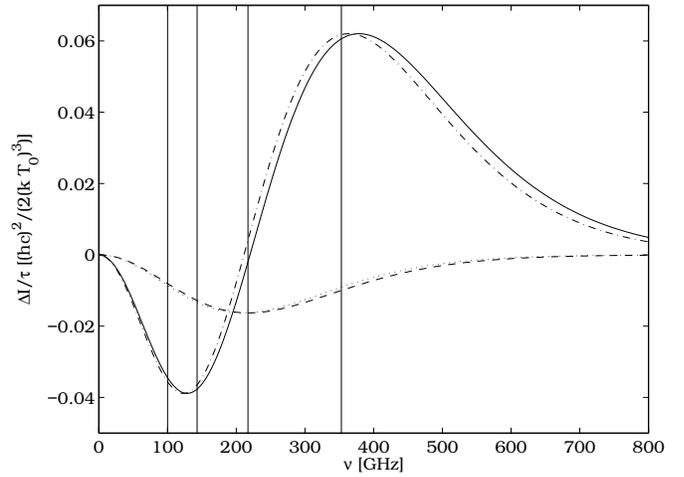,width=88mm}
\caption{Spectral signature of both the thermal SZ effect (solid line) 
and the kinetic effect (dashed line) for the standard cosmological temperature 
law ($a=0$). 
The dash-dotted and the dotted lines show the corresponding curves for 
$a=0.1$ and $z=0.5$. 
The {\it Planck} frequencies used in this study are indicated. 
The following cluster parameters were used: 
$kT_e=5$ keV,  $v_r = + 1000$ km s$^{-1}$. 
}
\label{fig1}
\end{figure}

\begin{table}
 \def\hf{\hfill}
 \halign
{#&# &# &# &# &#\cr
 \hline
 \noalign{\smallskip}
Central frequency (GHz)\hf			&100	&143	&217	&353\cr
Bandwidth (GHz) \hf		&33	&47	&72	&116\cr
Angular resolution (arcmin)\hf 	&9.2	&7.1	&5.0	&5.0\cr
Effective beam area (arcmin$^2$)\hf &95.9 	&57.1	&28.3	&28.3\cr
\hf     &\hf     \cr
Random noise (mJy/beam)\hf 	&14.0	&10.2	&14.3	&27.0\cr
Cosmic Infrared Background (mJy/beam)\hf &10.5 &15.6 &23.7 &81.9\cr
Unresolved radiosources (mJy/beam)\hf  &26.0 &15.5 &8.6 &\, -\cr
Total uncertainty (mJy/beam)\hf			&31.4 &24.3	&29.0	&86.3\cr
 \noalign{\smallskip}
 \hline
 \noalign{\smallskip}
}
\caption{Four frequency channels of the {\it Planck} high-frequency 
instrument relevant to this paper. 
The effective beam area has been calculated assuming a Gaussian beam.
The levels of contamination due to the Cosmic Infrared Background and 
to unresolved radiosources are taken from \cite{2004astro.ph..2571A}. 
}
\label{table1}
\end{table}
Another way to probe the temperature of the CMB  
is provided by the Sunyaev-Zeldovich (SZ) effect 
(\citealt{1969Ap&SS...4..301Z}, 
\citealt{1972CoASP...4..173S}). 
The CMB spectrum is distorted due to interaction with intervening hot electrons. 
Massive clusters of galaxies contain significant amounts of hot gas that can produce
a measurable SZ signal 
(see Fig.~\ref{fig1}).
Both the decrement at centimeter wavelengths and the increment in the millimeter range  
have been detected and large dedicated SZ surveys are planned (see 
\citealt{1995ARA&A..33..541R}, 
\citealt{1999PhR...310...97B}, 
\citealt{2002ARA&A..40..643C}  
for reviews). 
Using the standard scaling $\nu = \nu_0 (1+z)$ and the
temperature-redshift relation given in eq.~(\ref{eq1}), the dimensionless 
frequency $x\equiv h\nu/kT$ 
depends on $z$, and we have  
\begin{equation}
x\equiv \frac{h\nu}{kT} = \frac{h\nu_0}{kT_0} (1+z)^a \, ,
\label{eq2}
\end{equation}
where $h$ is the Planck constant and $k$ is the Boltzmann constant. 
The redshift independence of the SZ effect therefore occurs only 
when the standard temperature law is assumed. 

Two methods can provide information on the CMB temperature at the redshift
of the intervening cluster 
(\citealt{1978Ap&SS..59..223F}, 
\citealt{1980ApJ...241..858R}). 
The first one 
involves a precise measurement of the null point of the SZ signal, 
located around 218~GHz. 
The precise frequency depends both on the temperature of the hot gas and its redshift 
(see Sect.~\ref{sect2}). It also depends on the peculiar velocity of
the cluster.
The second method involves measurement of the SZ signal at several  
frequencies; 
the temperature of the CMB can be derived from the ratio of the observed SZ flux densities 
in different bands, which for some frequencies depends weakly on the properties of the 
intra\-cluster gas (see Sect.~\ref{sect3}).  

In this paper, we investigate the possibility of applying these methods to upcoming
SZ observations such as those planned in the {\it Planck} mission 
(see Table~1, and 
\citealt{2003NewAR..47.1017L}, 
\citealt{2004AdSpR..34..491T}, 
\citealt{2004AIPC..703..401M}  
about this mission). 

\section{The SZ effect}

The {\it thermal SZ effect} can be described in terms of the Compton $y$ parameter: 
\begin{equation}
y= \sigma_T \int n_e \left( \frac{kT_e}{m_ec^2}\right) dl 
\end{equation}
where 
$\sigma_T=6.65\times 10^{-25}$~cm$^2$ is the Thomson cross section, 
$n_e$ is the number density of electrons, 
$T_e$ is the temperature of the intracluster gas, 
$m_e$ is the electron mass, and 
$c$ is the speed of light. 
The $y$ parameter is a measure of the integrated pressure
of the intracluster gas along the line-of-sight. 

Due to inverse Compton scattering off the hot electrons, the
spectrum of the CMB is distorted. 
In the non-relativistic limit 
(if the velocity of the hot electrons is negligible with respect to $c$), 
the change in the CMB flux density is
\begin{equation}
S_\nu(x) = S_\nu^{\rm CMB}(x) Q(x) Y 
\end{equation}
where
$S_\nu^{\rm CMB} = 2(kT)^3 x^3/((hc)^2(e^x-1))$ is the unperturbed CMB flux density. 
The spectral distortion of the incident CMB spectrum is described by the 
function 
\begin{equation}
Q(x)= \frac{x e^{x}}{e^{x}-1}\left(
\frac{x}{\tanh{(x/2)}}-4
\right) \,.
\end{equation}
$Y$ is the integrated Compton parameter over the projected area of the cluster 
on the sky: 
\begin{equation}
Y = D_A^{-2}\int y dA \, ,
\end{equation}
where $D_A$ is the angular diameter distance. 

For high electron temperatures the relativistic effect has to be taken
into account 
(e.g., \citealt{1979ApJ...232..348W},  	
\citealt{1981Ap&SS..77..529F},   	
\citealt{1997ApJ...481L..55R}).  	
Useful numerical tables and analytic fitting formulae have been provided by 
several authors (see \citealt{2004A&A...417..827I} 
and references therein). 

The {\it kinetic SZ effect} is due to the motion of the cluster with respect
to the CMB frame: 
\begin{equation}
\frac{\Delta T}{T} = -\tau \frac{v_r}{c} \, , 
\end{equation}
with the convention that $v_r>0$ for a receding cluster. 

\section{Measurement of the cross-over frequency}\label{sect2}
\begin{figure}
\psfig{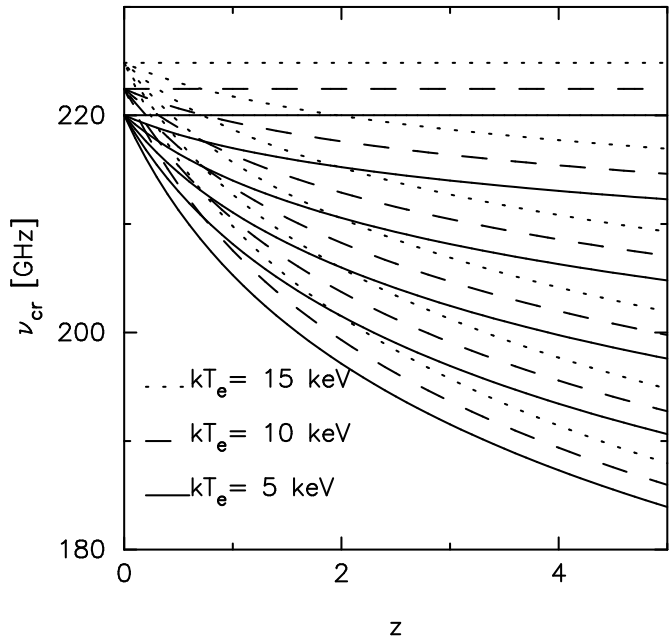}
\caption{
Variation of the cross-over frequency as a function of 
the redshift of the intervening cluster 
for different cluster temperatures and different $T-z$ relations 
(the parameter $a$ varies from 0 to 0.1 by 0.02 from top to bottom). 
Here the peculiar velocity of the clusters has been taken as equal to 0. 
}
\label{fig2}
\end{figure}

The location of the null point of the SZ signal $\nu_{\rm cr}$
can give information on the temperature of the CMB at the redshift
of the intervening cluster if it doesn't obey the standard law. 
From eq.~(\ref{eq2}) we have $\nu_{\rm cr} = x_{\rm cr}\frac{kT_0}{h}(1+z)^{-a}$. 
Sazanov \& Sunyaev (\citeyear{1998ApJ...508....1S}, 
\citeyear{1998AstL...24..553S}) 		    
provide a simple approximation for the non-dimensional cross-over
frequency $x_{\rm cr}$:
\begin{equation}
x_{\rm cr} = 3.830 \left( 1 + 0.30 {{v_r}\over{c}}{{m_e c^2}\over{kT_e}} 
+ 1.1 {{kT_e}\over{m_ec^2}} + 1.5{{v_r}\over{c}} \right) 
\label{eq9}
\end{equation}
where $v_r$ is the velocity of the cluster along the line of sight and $T_e$ is 
the temperature of the intracluster gas. 
For $v_r = 0$ one recovers the 
approximation given by \cite{1981Ap&SS..77..529F}. 
\cite{1998ApJ...502....7I} 
provide a second-order fit to the numerical result, 
$x_{\rm cr}= 3.83 \left( 1 + 1.1674 \theta_e - 0.8533\theta_e^2 \right)$\, , 
valid for $0< kT_e < 50$~keV and more accurate at high $T_e$ than 
eq.~(\ref{eq9}), and where $\theta_e= kT_e/m_e c^2$. In the following
we use that approximation in the figures where $v_r = 0$.

Figure~{\ref{fig2}} shows the variation of the cross-over frequency as a
function of the redshift of the intervening cluster
for different cluster temperatures and different $T-z$ relations. 
Here we first consider $v_r=0$. 
For $a=0$, one recovers the known result that $\nu_{cr}$ is constant
with redshift and increases with increasing $T_e$.
For different values of $a$, the curves are split and $\nu_{\rm cr}$
decreases with increasing $a$. There is thus a parameter degeneracy 
between $T_e$ and $a$. 

\begin{figure}
\psfig{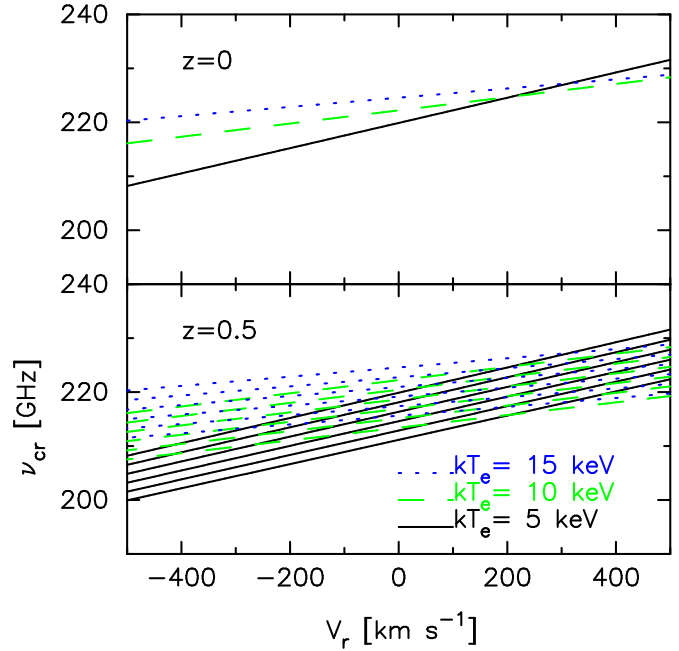}
\caption{
Variation of the cross-over frequency as a function of 
the peculiar velocity of the intervening cluster 
for different cluster temperatures and different $T-z$ relations 
(the parameter $a$ varies from 0 to 0.1 by 0.02 from top to bottom). 
}
\label{fig3}
\end{figure}

Figure~{\ref{fig3}} illustrates the dependence on the peculiar
velocity of the cluster, $v_r$, introducing a further degeneracy. 

Those parameter degeneracies and the large bandwidths of current bolometers 
make it difficult to apply this method in practice to constrain the CMB
temperature. 

\section{Measurements at different frequencies}\label{sect3}
A more promising method 
consists of using ratios of observed flux densities in several frequency
bands. 
It has recently been applied sucessfully to determine the
CMB temperature at the redshift of the Coma cluster (at $z=0.0231$)
and of Abell~2163 at $z= 0.203$ by 
\cite{2002ApJ...580L.101B} 
who used observations at 32, 143 and 272~GHz.

\begin{figure}
\psfig{file=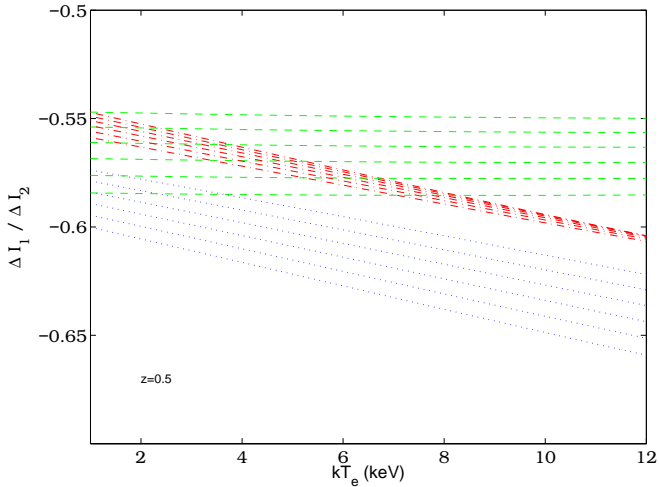,width=88mm}
\caption{
Variation of the flux density ratio as a function of 
the cluster temperature 
for clusters at $z=0.5$ and 
different $z-T$ relations.  
The parameter $a$ varies from 0 to 0.1 by 0.02 from bottom to top. 
Here the peculiar velocities have been set to zero. 
The dashed lines show the 100/143~GHz flux density ratio, multiplied by $-0.6$ for clarity. 
The dash-dotted lines show the 143/353~GHz ratio. 
The dotted lines show the 100/353~GHz ratio. 
Clearly, the 100/143~GHz ratio is most sensitive to $a$ and varies least with  
the cluster temperatures. 
}
\label{fig4}
\end{figure}

\begin{figure}
\psfig{file=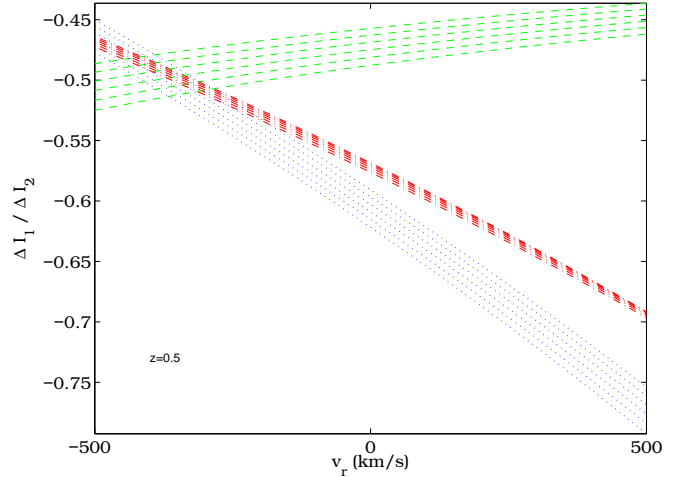,width=88mm}
\caption{
Same as Fig.~{\ref{fig4}} but as a function of the peculiar velocity
of the cluster. 
The 100/143~GHz flux density ratios have been multiplied by $0.5$ for clarity. 
Here $kT_e = 5$~keV. 
The 143/353~GHz flux density ratio is most sensitive to the peculiar velocities. 
}
\label{fig5}
\end{figure}

We have calculated the flux density of the thermal SZ effect using the analytic fitting formulae
given by \cite{1998ApJ...502....7I} 
valid in the relativistic case. 
We have added the contribution of the kinetic effect. 
The flux density ratio depends weakly on the cluster temperature for frequencies
between 32~GHz and $\sim170$~GHz.  
If the second measurement ($\nu_2$) is done at a higher frequency, 
the flux density ratio decreases with increasing 
$T_e$, more strongly as $\nu_2$ increases. 
For different values of the $a$ parameter, the curves are split. 
The magnitude of the split increases with redshift. 
Figure~\ref{fig4} shows  
the variation of the flux density ratios with the cluster temperature,  
Fig.~\ref{fig5} the variation with the peculiar velocity of the cluster. 
Observations with new X-ray satellites (XMM, Chandra) make it possible
to measure $T_e$ with good accuracy. 
Note, however, that the flux density ratios depend only weakly on
the cluster temperature. 
To constrain the parameter $a$, it would be necessary to constrain 
the peculiar velocity of a cluster independently. 
Alternatively, it is possible to constrain $a$ by observing a 
large number of clusters with known 
gas temperatures, as will be shown in the next section. 

\section{Simulating a large SZ survey}

We seek to simulate the yield of a {\it Planck}-like SZ survey to test how
effectively our parametrization of the $T-z$ relation can be constrained.
One way to do this is to simulate sky maps 
based on $N$-body simulations 
with SZ signals as well
as random and systematic noise in the relevant frequency bands and then
use a suitable cluster extraction algorithm (see, for example,
\citealt{2005MNRAS.360...41G}). 
Here we take the simpler approach of
directly simulating a cluster catalog from the redshift and mass
distribution of dark matter halos and computing the SZ signals assuming that 
the clusters are unresolved. This method  is justified for this study as
it gives approximately the same number of recovered clusters as more
general methods. 

\subsection{The model}

We have used a fiducial $\Lambda$CDM cosmology with 
$\Omega_0=0.3$, 
$\lambda_0=0.7$,
$H_0=100 h$~km~s$^{-1}$Mpc$^{-1}$, 
$h=0.7$,
$\sigma_8=0.9$, consistent with the first-year {\it WMAP} results 
(\citealt{2003ApJS..148..175S}). 	    
We have used the \cite{1999MNRAS.308..119S} 
cluster mass function, which provides an excellent  
match to the results from large numerical simulations, and has the advantage
of having a theoretical justification in terms of the ellipsoidal collapse 
of clusters
(\citealt{2001MNRAS.323....1S}).  
The mass function can be expressed as (\citealt{2002MNRAS.336..112M}): 
\begin{equation}
n(M,z) dM 
=  
A\left( 
1 + {{1}\over{\nu'^{2q}}}
\right)
\sqrt{{2}\over{\pi}} 
{{\rho_0}\over{M}} 
{{d\nu'}\over{dM}}
e^{(
-{{\nu'^2}\over{2}}
)  
}dM
\label{eq-st}
\end{equation}
where 
$\rho_0$ is the comoving background density,
$\nu'=\sqrt{\alpha}\nu$,
$\alpha=0.707$,
$A\simeq0.322$, 
$q=0.3$, 
$\nu = \nu(M,z) = \delta_c/(D(z)\sigma(M))$, 
$\delta_c$ is the linear overdensity of
a perturbation at the time of collapse and virialization;
$\delta_c$ varies weakly with the cosmological parameters and
has a value of $\delta_c\simeq1.69$.
$\sigma(M,z)$ is the variance of the fluctuation spectrum on
a mass scale $M = {{4}\over{3}}\pi R^3 \rho_0$. 

The number density of clusters above a certain mass
per unit steradian
per redshift interval is
\begin{equation}
{{d^2N(z)}\over{d\Omega dz}} = 
{{d^2V_c}\over{d\Omega dz}}
\int_{M_{\rm lim(z)}}^{+\infty} n(M,z) dM
\label{eq-d2n}
\end{equation}
where the comoving volume is
\begin{equation}
{{d^2V_c}\over{d\Omega dz}} = c{{D^2_A(z) (1 + z)^2}\over{H(z)}} \, ,
\end{equation}
where $D_A$ is the angular diameter distance and $H(z)$ is
the Hubble parameter.

\begin{figure*}
\psfig{file=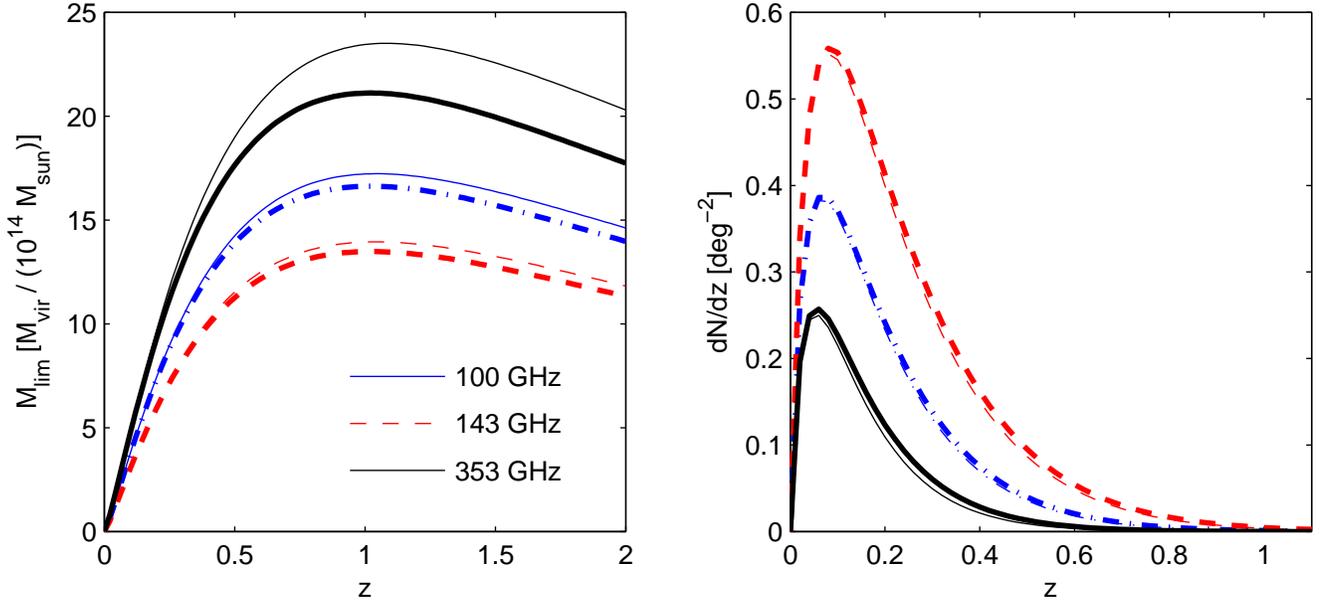,width=176mm}
\caption{
{\it Left:} Limiting mass as a function of redshift at three frequencies 
for observations at 3~$\sigma$ level using {\it Planck}, 
taking into account both the random instrumental noise and the levels of contamination 
due to the Cosmic Infrared Background
and to unresolved radiosources given in Table~1.
The thin lines correspond to the relativistic calculation and the thick lines to 
the non-relativistic one. 
} 
{\it Right:} Number density of clusters per redshift interval and per square degree
on the sky as a function of redshift, above the limiting mass.
\label{fig6}
\end{figure*}

A mass-temperature relation is needed  
in order to assign a gas temperature to a cluster. 
If the gas is in hydrostatic equilibrium with the gravitational potential
of the cluster, then: 
\begin{equation}
kT_e = \vline\frac{1}{
d\ln\rho_{\rm gas(r)}/d\ln r\, \vline_{r_{\rm vir}}
}\vline \, 
\mu m_P \frac{GM_{\rm vir}}{r_{\rm vir}}
\end{equation}
where $\rho_{\rm gas}$ is the gas density, 
$r_{\rm vir}$ is the virial radius, and 
$\mu=4/(5X+3)$ is the mean molecular weight, 
$X$ is the hydrogen mass fraction and
$m_P$ is the mass of the proton. 

We have adopted the \cite{1997ApJ...490..493N} 
profile for the gas:
\begin{equation}
\frac{d\ln\rho_{\rm gas}}{d\ln r}\vline\,_{r_{\rm vir}} = -\frac{1+3c}{1+c}
\label{eq-navarro}
\end{equation}
with
\begin{equation}
c= 6\left(
\frac{M_{\rm vir}}{10^{14} h^{-1} {\rm M}_{\odot}}
\right)^{-1/5} \, .
\end{equation}
Taking a constant $c=5$ yields an error in $\frac{d\ln\rho_{\rm gas}}{d\ln r}\vline\,_{r_{\rm vir}}$ in eq.~(\ref{eq-navarro}) of less than 5\% 
for the range of masses considered. 

Introducing $\Delta_c$, the average density within the cluster relative
to the critical density of the background at the redshift of the cluster, 
one obtains: 
\begin{equation}
\begin{array}{ccl}
\frac{kT_e}{\rm keV} 
	    &= 2 |\frac{7.75}{
d\ln\rho_{\rm gas(r)}/d\ln r|_{r_{\rm vir}}
}|  \\
	    & \times  \left(
\frac{6.8}{5X+3}
\right) 
\left(
\frac{M_{\rm vir}}{10^{15} h^{-1} M_{\odot}}
\right)^{2/3}\\
	    & \times 
(1+z) \left[
\frac{\Omega_0}{\Omega(z)}
\right]^{1/3} 
\left(
\frac{\Delta_c}{178}
\right)^{1/3} \, .
\end{array}
\label{eq-kte}
\end{equation}
We have calculated $\Delta_c$ from $\Delta_{\rm vir}=\rho/\rho_{\rm bg}$, 
taken from the fits given by 
\cite{1996ApJ...469..480K}: 
\begin{equation}
\Delta_{\rm vir} \approx 18\pi^2 (1 + 0.4093 w_{\rm vir}^{0.9052})\, ,
\end{equation}
where 
$w_{\rm vir}=1/\Omega_{\rm vir}-1$, and $\Omega_{\rm vir}$ is the density parameter at virialisation; 
$\Omega_{\rm vir}= {{\Omega_0 (1+z_{\rm vir})^3}\over{\Omega_0 (1+z_{\rm vir})^3 + \lambda_0}}$
for a flat $\Lambda$CDM model. 

We have added a spread of 30\% to the mass-temperature relation. 

As for the peculiar velocities of the clusters, we have drawn them from a Gaussian
distribution with zero mean.
\cite{2001MNRAS.322..901S} 
provide a fit for the dispersion as a function of the mass of the cluster:
\begin{equation}
\label{eqvpec}
\sigma_{\rm halo} (M)= \frac{\sigma_{\rm fit}}{1+(R/R_{\rm fit})^\eta}
\end{equation}
with 
$R=(3M/4\pi\rho_0)^{1/3}$, 
$\sigma_{\rm fit}= 414.7$~km~s$^{-1}$, 
$R_{\rm fit}=34.67 h^{-1}$~Mpc, and 
$\eta=0.85$ for a $\Lambda$CDM model with the same parameters as ours. 
$\sigma^2_{\rm halo}$ is the dispersion of the Maxwellian distribution
(three-dimensional peculiar velocity distribution). For the peculiar velocity along
the line-of-sight, we thus took
$\sigma^2_{\rm pec}= \sigma^2_{\rm halo}/3$. 

\begin{figure*}
\psfig{file=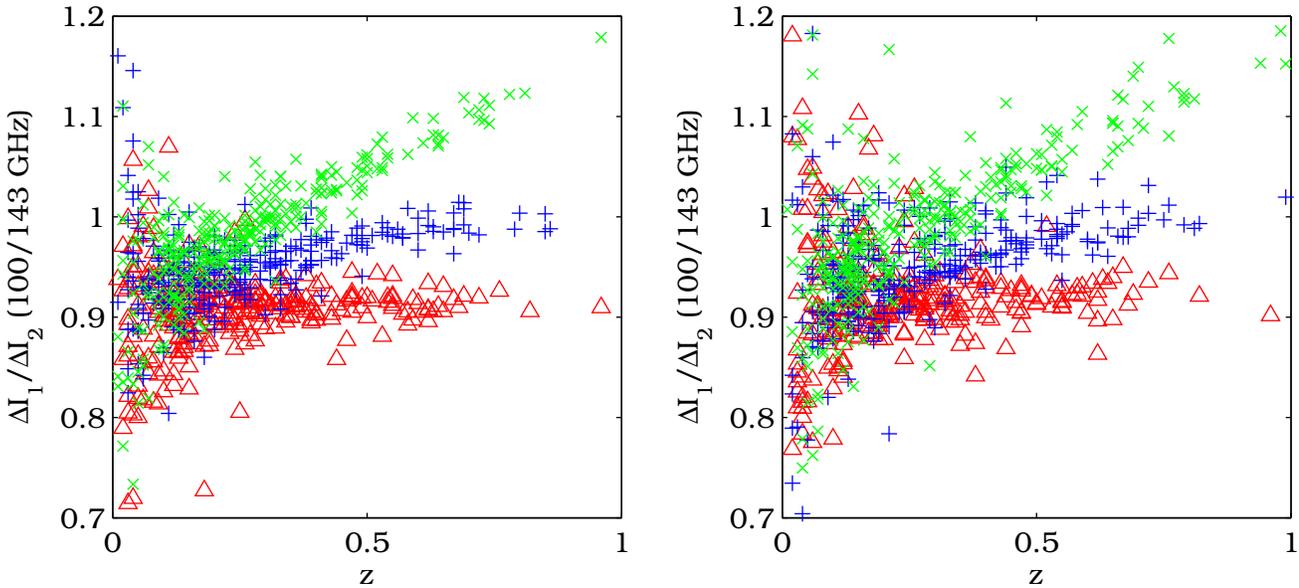,width=176mm}
\caption{
Flux density ratio versus redshift for simulations with different values of the parameter
$a$. 
For clarity, only instrumental noise has been included.
The triangles correspond to $a=0$, 
the plusses to $a=0.1$, 
the crosses to $a=0.2$. 
{\it Left:} In the absence of peculiar motions ($v_{r}=0$). 
The distributions are clearly different. 
{\it Right:} Including the peculiar motions. The spread in the flux density ratio is much
larger than when $v_r=0$, and the different $a$ distributions are mixed at low redshifts. 
}
\label{fig7}
\end{figure*}

\subsection{The limiting mass}\label{subsec-limmass}
The integration of eq.~(\ref{eq-d2n}) is to be done from a limiting mass 
which is redshift-dependent.  
The majority of the clusters will be unresolved at the resolution of the
{\it Planck} instruments (\citealt{2001MNRAS.325..835K}). 

The SZ flux  can be expressed as 
\begin{equation}
\begin{array}{ccl}
S_\nu  & =  
       & 2.29\times10^4\frac{x^3}{e^x-1} Q(x)\times 1.70\times 10^{-2}
                     h \left(\frac{f_{\rm ICM}}{0.1}\right)\\
       &\times &\left( \frac{1+X}{1.76}\right) 
               \left( \frac{D_A}{100 h^{-1} {\rm pc}} \right)^{-2}  
               \left(  \frac{kT_e}{\rm keV} \right)  \, {\rm mJy}\, , \\
\end{array}
\end{equation}
where 
$f_{\rm ICM}$ is the gas mass fraction that we will take constant and equal to 0.1 
and $kT_e$ is taken from eq.~(\ref{eq-kte}) 
(\citealt{2001ApJ...550..547F}).  

We have calculated the limiting mass for $\it Planck$ observations 
at a detection level of 3~$\sigma$, where $\sigma$ is the total uncertainty
given in Table~1. 
The error estimates on the flux densities are discussed in Sect.~\ref{subsect_errorestim}. 

\subsection{The SZ catalog}

We have distributed clusters in mass and redshift 
according to eq.~(\ref{eq-st}) and eq.~(\ref{eq-d2n}), 
above the limiting mass discussed above 
(see Fig.~\ref{fig6}). 
After excluding the nearby clusters at redshifts below 0.05, 
we obtain a catalog of $\sim 1200$ clusters.  

We have then calculated the flux density ratios for various frequency pairs and for different
values of the parameter $a$ in the cosmological redshift-temperature law. 
We have used the bandwidths of the {\it Planck} instruments and calculated the 
observed SZ flux density assuming a top-hat response of the filters  
($R=1$ for $\nu_c - \Delta \nu/2<\nu< \nu_c + \Delta \nu/2$ and $R=0$ elsewhere 
(see Table~1)). 

\subsection{Error estimates}\label{subsect_errorestim}
Errors are simulated through use of the noise levels quoted in Table~1. 
As for the uncertainties on the flux density ratios, we have 
propagated the uncertainties on the individual flux densities. 

The random noise and the contribution from the cosmic
infrared background and from unresolved radiosources have been taken from 
\cite{2004astro.ph..2571A}  
and added in quadrature.  
Dust emission has been neglected as the levels of contamination are orders of 
magnitude lower than for the other contaminants mentioned above.

Another contaminating source is the primary temperature fluctuations of the CMB, which are 
important for observations with a large beam such as that of {\it Planck}. 
The primordial anisotropies have the same spectral signature as the kinetic SZ effect. 
Promising techniques to separate the primary CMB fluctuations from the kinetic SZ effect 
have been developed, for example by taking advantage of the spatial correlation between
the thermal and the kinetic effects (see \citealt{2004A&A...420...49F} 
and references therein). 
In the present study, we have assumed that the contamination from the CMB could be removed.
Observations with a smaller telescope beam close to the average size of the cluster 
(around 1 arcmin) wouldn't suffer significantly from  
contamination from the CMB. It has also been shown that such observations would provide
an optimum survey yield for the detection of clusters (\citealt{2004astro.ph.10392B}). 

Although we assume precise and unbiased knowledge of the
electron temperatures in the catalog, 
it should be pointed out that the results are not very sensitive
to the cluster temperatures, as shown in Fig.~4. Adding an uncertainty of 
up to 20\% on the cluster temperatures doesn't affect the final result
significantly. 
Errors due to peculiar velocities have to be modelled more
carefully, however; not only because of the limited precision with which
these
can be extracted from the SZ data, but also because $\Delta I_1/\Delta
I_2$ is more
sensitive to $v_r$ than to $kT_e$.

We use a separate frequency ratio to fit to
$v_r$ for each individual cluster. Error bars on each peculiar
velocity are converted into error bars in the ratio used
for the CMB temperature by simply calculating this ratio at the
endpoints of the $68 \%$ confidence interval of the peculiar velocities.  These 
systematic errrors are then added in quadrature to the other errors.

\subsection{Analysis}
As a first step, we have compared simulated SZ observations of catalogs 
with different $a$. 
Figure~\ref{fig7} shows the flux density ratio for the frequency pair 
100/143~GHz  
as a function of redshift in two cases: first, in a simulated catalog
where all clusters have zero peculiar velocity (left), and then 
where peculiar velocities have been assigned according to eq.~(\ref{eqvpec}). 
Clearly, the introduction of peculiar velocities confuses the picture and makes
it more difficult to distinguish the distributions with different $a$. 
In order to compare quantitatively the distributions with the inclusion of 
peculiar velocities, 
we have run a 
Kolmogorov-Smirnov test (Fig.~\ref{fig8ks}). 
We see clearly that the distributions are statistically different for a sufficiently
large $a$. 

This first test is encouraging. In practice, however, one would like to be able
to determine the value of the $a$ parameter from {\it one} single simulation, since
we have only one observable universe at our disposal! 
To do this, we have devised the following iterative method: 

\begin{figure}
\psfig{file=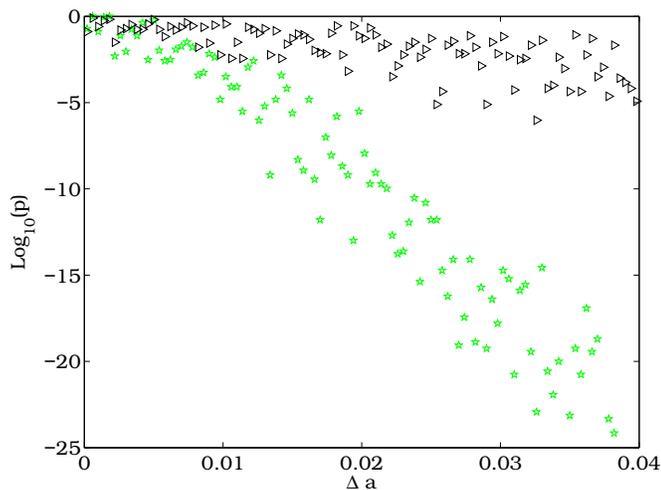,width=88mm}
\caption{
Kolmogorov-Smirnov test: 
The significance level $p$ is shown as a function of $a$. 
$p$ is the probability that the two data sets are drawn from the same distribution. 
Stars show the results for the 100/143~GHz frequency pair, 
triangles for the 143/353~GHz pair. 
The 100/143~GHz pair appears as the better one, 
as the probability that the $a=0$ distribution
be the same as a one with $a\neq0$ drops most rapidly with increasing $a$.
Each data set contains 1000 clusters. 
}
\label{fig8ks}
\end{figure}
\begin{figure*}
\psfig{file=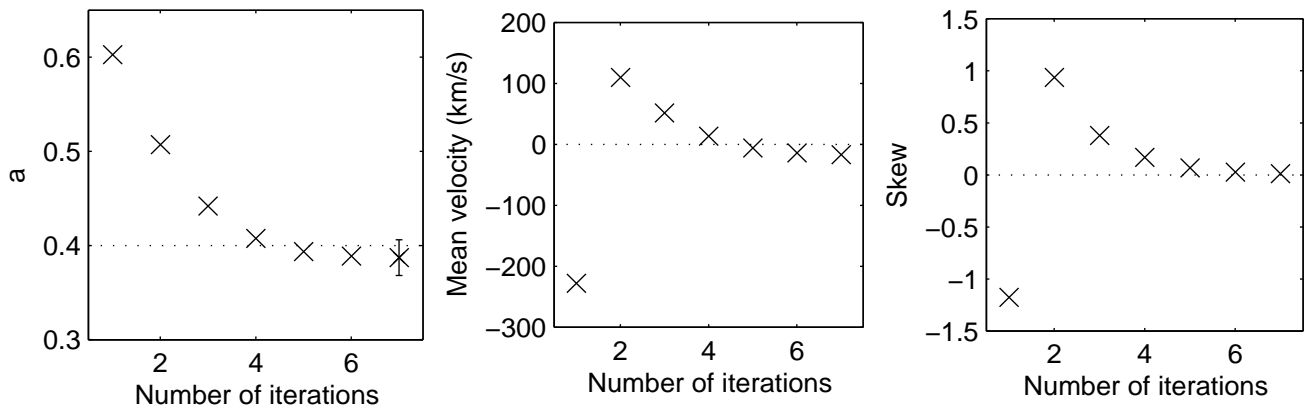,width=176mm}
\label{figconvergence}
\caption{ 
Convergence of the method. 
Here a simulation was $a=0.4$ was performed. 
The method clearly converges toward that value of $a$, as can also be seen
from several diagnositics: 
the mean value of the peculiar velocity distribution and 
the skewness converge toward zero as $a$ tends toward its
original value. 
}
\label{fig9convergence}
\end{figure*}

\begin{enumerate}
\item{First assuming $a=0$, we do a $\chi^2$ analysis 
to fit the peculiar velocities
to the simulated flux density ratios at two frequencies, namely 143 and 353 GHz, which
is the frequency pair which is most sensitive to peculiar velocities. 
We do this for each cluster. 
}
\item{Using those velocities, we then do a $\chi^2$ analysis to fit the whole catalog
to $a$, now using the frequency pair 100/143~GHz, which is most sensitive to $a$.}
\item{Taking the parameter $a$ obtained at step~2, we iterate 
and fit to the peculiar velocities, and
then fit to a new value of $a$. }
\end{enumerate}

We iterate until the method converges.  
Figure~\ref{fig9convergence} shows clearly that it does: 
Here we have deliberately used a large $a$: 0.4. 
After a few iterations, the method converges.
We have also examined the distribution of the peculiar velocities at each iteration. 
At the first iteration (when $a$ is taken equal to zero), the distribution of 
peculiar velocities is clearly non-Gaussian, as indicated by its skewness.  
With each integration, the distribution of velocities 
converges toward the original one; the parameter $a$ also tends toward
its original value.  
In order to test the robustness of the method, we repeated 200 times a simulation 
with $a=0$. The distribution of the inferred $a$ parameters 
follows a Gaussian, with a dispersion around $10^{-2}$.

\section{Conclusion}
We have examined the possibility to constrain the cosmological temperature-redshift
law from multifrequency SZ observations. 
The advantage of using flux density ratios is to eliminate a possible dependence on 
cluster parameters, in particular the optical depth and, to some extent, 
the cluster temperature. 
After several tests, we have selected the 100/143~GHz frequency pair as 
particularly sensitive to our parametrization of the $T-z$ law, 
and the 143/353~GHz pair as sensitive to the cluster peculiar velocities. 
The iterative method that we have devised makes it possible to recover 
the parameter $a$ of the $T-z$ law from SZ flux density ratios of a large 
number of clusters. Our simulated catalog contains $\sim 1200$ clusters. 
We have assumed a $\Lambda$CDM cosmology, a full-sky survey and detection at more than 
three times the expected sensitivity of the {\it Planck} instrument
taking into acccount contamination by the Cosmic Infrared Background and radiosources. 
Note, however, that the method is independent of the cosmological parameters
and could be applied to any catalog of clusters for which redshifts and, 
to some extent, 
temperatures are known.  

Most cosmological models predict a linear variation of the
CMB temperature with redshift. 
Sensitive SZ observations appear as a potentially useful tool to test the standard law. 
The discovery of an alternative law would have profound implications 
on the cosmological model, implying creation of energy in a manner that
would still maintain the black-body shape of the CMB spectrum at 
redshift zero.

\begin{acknowledgements}
We are grateful to John Black for useful comments on the manuscript.
C.H. is grateful to John Bahcall for an invitation to visit the Institute
for Advanced Study in Princeton in September 2004 where part of this work 
was completed.
C.H. acknowledges financial support from the Swedish Research Council
{\it Vetenskapsr\aa det}.   
\end{acknowledgements}

\bibliography{Biblio/horellou_ms3090-biblio} 

\end{document}